\documentclass[prodmode,hillsideplop]{acmlarge}
\usepackage{flafter}
\usepackage{hyperref}
\usepackage[hyphenbreaks]{breakurl}

\usepackage{todonotes}
\usepackage{xcolor}
\presetkeys%
    {todonotes}%
    {inline,backgroundcolor=yellow}{}
\acmVolume{22}
\acmYear{2021}
\acmMonth{06}

\usepackage[ruled]{algorithm2e}
\SetAlFnt{\algofont}
\SetAlCapFnt{\algofont}
\SetAlCapNameFnt{\algofont}
\SetAlCapHSkip{0pt}
\IncMargin{-\parindent}

\title{More Software Analytics Patterns:\\ Broad-Spectrum Diagnostic and Embedded Improvements}
\author{
  Duarte Oliveira \affil{Faculty of Engineering, University of Porto. Porto, Portugal}\ \\
  João Fidalgo \affil{Faculty of Engineering, University of Porto. Porto, Portugal}\ \\
  Joelma Choma \affil{National Institute for Space Research - INPE. Brazil}\ \\
  Eduardo Guerra \affil{Free University of Bozen-Bolzano. Bolzano, Italy}\ \\
  Filipe F. Correia \affil{Faculty of Engineering, University of Porto. INESC TEC. Porto, Portugal}
}

\appendixauthor{IF DIFFERENCE AUTHORS INCLUDE HERE, University of New Hampshire}

\begin{abstract}
Software analytics is a data-driven approach to decision making, which allows software practitioners to leverage valuable insights from data about software to achieve higher development process productivity and improve different aspects of software quality. In previous work, a set of patterns for adopting a lean software analytics process was identified through a literature review. This paper presents two patterns to add to the original set, forming a pattern language for adopting software analytics practices that aims to inform decision-making activities of software practitioners. The writing of these two patterns was informed by the solutions employed in the context of two case studies on software analytics practices, and the patterns were further validated by searching for their occurrence in the literature. The pattern \textsc{Broad-Spectrum Diagnostic} proposes to conduct more broad analysis based on common metrics when the team does not have the expertise to understand the kind of problems that software analytics can help to solve; and the pattern \textsc{Embedded Improvements} suggests adding improvement tasks as part of other routine activities. 

\end{abstract}



\acmformat{Oliveira et al. 2021. More Software Analytics Patterns: Broad-Spectrum Diagnostic and Embedded Improvements.}

\copyr{PLoP'21, SEPTEMBER 16--30, , {USA}. Copyright 2021 is held by the author(s). {HILLSIDE} 978-1-941652-03-9}

\begin{document}

\begin{bottomstuff}
Permission to make digital or hard copies of all or part of this work for personal or classroom use is granted without fee provided that copies are not made or distributed for profit or commercial advantage and that copies bear this notice and the full citation on the first page. To copy otherwise, to republish, to post on servers or to redistribute to lists, requires prior specific permission. A preliminary version of this paper was presented in a writers' workshop at the 28th Conference on Pattern Languages of Programs (PLoP).
\end{bottomstuff}

\maketitle

\section{Introduction}
The term \textit{analytics} refers to the extensive use of data, analysis, and systematic reasoning to leverage in-depth information about a given issue and support professionals' decisions-making process from different areas~\cite{davenport2009make}. For instance, approaches using analytics have been applied for some time in the marketing area to reach and better understand customers from different markets~\cite{davenport2010organizations}. Most recently, analytics have also been widely used in the domain of software, as a way to support practitioners (\textit{e.g.}, developers, testers, designers, and managers) make better decisions regarding their processes, products, and services~\cite{zhang2011software}.

\citeN{zhang2011software} coined the term \textit{software analytics} to refer to the use of analytics to leverage insightful and actionable information from software data and inform the decisions of software practitioners~\cite{zhang2011software}. By \textit{software data}, we mean the data generated as a result of developing a software system. Some of it may be recorded in repositories, such as version-control and issue-tracking systems, and includes artifacts such as source code, bug reports, or test execution reports. Other such data may be generated during the operation of the system and can be found in artifacts such as log files~\cite{storey2016lies}. The typical issues addressed by software analytics are related to software failures and defect prediction, software requirements, code quality, releases and code integration, project management, teamwork, software maintenance, and software evolution, among others~\cite{buse2010analytics,hassan2010software}.

Although widely adopted by software companies, software analytics is often not explored to its full potential. Some studies have lead to the proposal of analytics methods and tools, but few of them provide detail on how to adopt software analytics practices in real-world projects~\cite{zhang2013software,huijgens2018software,augustine2018deploying,snyder2018using}. Furthermore, many software practitioners are still reluctant to adopt software analytics practices~\cite{robbes2013software}. For different business reasons, and especially to meet customers' expectations, many teams focus on external quality aspects (the functionalities as perceived by end-users) and devote less time than what they would like to internal quality aspects (\textit{e.g.}, architecture, maintainability, flexibility, security).

In a previous article, we propose eight patterns for software analytics (SA)~\cite{choma2018patterns,choma2017patterns}, that will help a team to adopt software analytics practices in their projects. These patterns have later helped us to propose the Software Analytics Canvas (SA Canvas)~\cite{chomatowards}, with the goal of consolidating the patterns and making their application as practical as possible. The SA Canvas is designed to support planning and managing software analytics activities during software development. By evaluating the canvas in practice, we have identified new patterns.

In this paper, we present two patterns identified during a study of the SA Canvas in two software companies. The first is a large company with multiple projects and strictly-structured teams. Given its business domain, the company takes security and data privacy particularly seriously. Our study was conducted with a small team from a recent project involving four developers, of which three had no experience and no knowledge of software analytics. This study took place over three sprints, totaling 12 weeks.
The second company is a software start-up. Start-ups are characterized by rapid evolution, uncertainty about customer and market needs, lack of resources, and small teams~\cite{startupuncertainties}. The study's main objective at this start-up was to evaluate the use of SA Canvas to improve the internal quality of their product. The study involved two teams whose members had no experience in software analytics. The former had four developers mainly focused on back-end development. The latter involved five developers, three of them focused on back-end development, and two focused on front-end development. Both teams followed the Scrum framework and adopted the SA Canvas during five sprints, totaling 10 weeks.

The two patterns presented in this paper arose from the difficulties we observed when applying SA Canvas at these two software companies. They describe solutions for overcoming obstacles when adopting software analytics practices. The first pattern can make the team more aware of the kind of information it can generate through the use of metrics and help it to find the first issues to tackle through analytics. The second pattern can be used when acting upon the analytics' findings, proposing how to implement improvements with a low impact on other tasks.

\section{Software Analytics Patterns}\label{sec:patters_sa}

In this section, we first provide an overview of the SA patterns we have published in previous studies~\cite{choma2018patterns,choma2017patterns}. Then, we briefly introduce the two new patterns proposed in this paper, describing how they fit into the previous ones. An overview of the SA patterns showing how they relate to each other is depicted in Figure \ref{overview}. The blocks with black background represent those patterns for adopting software analytics that we have published in previous studies, and those with gray background the patterns that we are proposing in this article. The blocks with white background and dashed black borders represent outputs to be expected from applying the patterns. The questions included next to relationship arrows motivate the use of the target pattern.

\subsection{Pattern Language Overview}
As previously mentioned, the first eight SA patterns emerged from a literature review, in which we searched for best practices in experience reports to identify the typical issues addressed with software analytics. A summary of the eight patterns containing a brief description of each of them is presented below.

\begin{itemize}
\item[(1)] \textsc{What You Need to Know}: To solve the issues that the team want to improve in the system and/or the software development process, in a context where there is a large amount of software data that can inform the decisions of the team, the solution is to define the key issues that the development team wants to focus on, in order to improve the software throughout the project.  

\item[(2)] \textsc{Choose the Means}: To solve how to gather useful data regarding the issues that the team need to solve, in a context where a plethora of data is available, the solution is to define the most appropriate means, such as metrics, tools, techniques and other approaches for extracting data from software artifacts that will be useful in future decisions.

\begin{figure}[!ht]
    \centering
    \begin{minipage}{1.0\textwidth}
        \centering
        \includegraphics[width=1.0\linewidth]{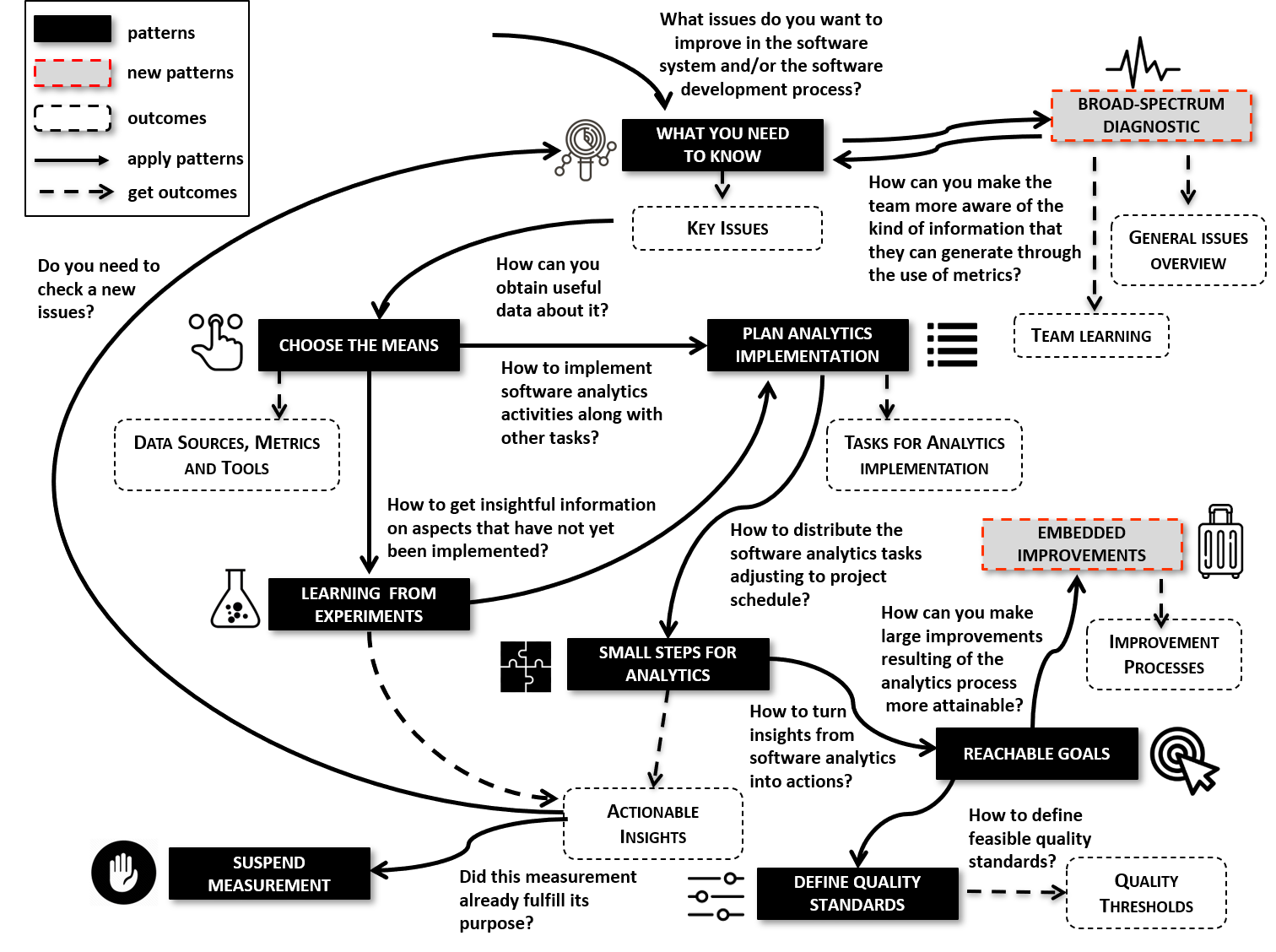}
        \caption{Overview of the patterns and their relationships.}
        \label{overview}
    \end{minipage}
\end{figure} 

\item[(3)] \textsc{Plan Analytics Implementation}: To solve how to schedule the software analytics activities fitting them to project roadmap along with other development tasks, in the context where the tasks directly related to the implementation of software features are the top priority, the solution is add tasks related to the software analytics in the backlog to be prioritized with the regular project tasks.

\item[(4)] \textsc{Small Steps for Analytics}: To solve how to schedule software analytics in a pace that it does not to overburden the team, in the context where much information at the same time can confuse and make the team lose focus, the solution is to adjust software analytics tasks within the team schedule by breaking down them at smaller portions to be carried out in multi-steps.

\item[(5)] \textsc{Reachable Goals}: To solve how to turn software analytics findings into actionable insights to improve software aspects, in a context where to perform all improvements based on the analytics automated feedback might lead the team to act without focus, the solution is to take actionable insights from the software analytics findings, and from them, settle reachable goal adjusting the action steps.

\item[(6)] \textsc{Learning from Experiments}: 
To solve how to obtain information to make informed decisions about software issues on some aspect we have not yet implemented or we need to redesign, in a context where the team has nowhere yet to collect and analyze data to support their decisions, the solution is to create an alternative solution and perform an experiment collecting data that allow the comparison with the current solution.

\item[(7)] \textsc{Define Quality Standards}: 
To solve how to achieve and maintain a good level of quality for important software aspects, in the context where the improvements can be made incrementally, the solution is to define quality standards and then establish minimal or maximum thresholds for any software aspect that the team intends to monitor.

\item[(8)] \textsc{Suspend Measurement}: 
To solve if an issue still need to be  continually monitored after some initial measurements, in a context where the team does not yet have a monitoring system, or the current system is overloaded with other issues, the solution is to put on standby the measurements that already fulfilled their initial goal, are costly to be continuously monitored, or that do not represent a value to the team at that moment.
\end{itemize}

According to the proposed patterns, the first step towards adopting software analytics practices is to define \textsc{What You Need to Know}. After that, with the purpose to answer the raised issues, the team needs to \textsc{Choose the Means} that will be used to data gathering and analysis. \textsc{Learning from Experiments} can be a way of testing a particular solution that the team is not sure if it is the best way from a practical standpoint. During the \textsc{Software Analytics Planning}, the team plans the analytics activities and prioritizes the tasks in their to-do list along with other development tasks. Because analytics activities can be time-consuming, the team do not have to be deployed them at once. Then, the team can set \textsc{Small Steps for Analytics}, according to delivery schedule. Based on actionable insights, the team needs to define \textsc{Reachable Goals} to incorporate the improvements in the software or in its development process. Towards continuous improvement, the team will \textsc{Define Quality Standards} to guide their improvement actions. The team can apply the pattern \textsc{Suspend Measurement} when measurements no longer make sense or when they have other priorities at the moment. 

\subsection{The new Software Analytics patterns}

From the difficulties we observed at two software companies when trying to implement software analytics in practice, we identified two new patterns that we now add to our pattern language, and represent in Figure \ref{overview} as the two gray blocks with dashed red borders.

The \textsc{Broad-Spectrum Diagnostic} pattern proposes to conduct more broad analysis based on common metrics when the team does not have the expertise to understand the kind of problems that software analytics can help to solve, contributing to team learning about existing issues overview and awareness about the analytics process.
As shown in Figure \ref{overview}, there are two arrows that connect this pattern to the \textsc{What You Need to Know} pattern. The upper one represents the team's shift in approach when they do not know where to start, or aren't aware of existing issues, or are not yet convinced of the need for software analysis. The bottom arrow refers to the move the team makes when it already has a \textsc{Broad-Spectrum Diagnostic} and now the team can move forward using the \textsc{What You Need to Know} pattern to appoint the most pressing issues to solve.

The \textsc{Embedded Improvements} pattern suggests adding improvement tasks as part of the team routine to guarantee a continuous improvement process. That pattern is an alternative to implement \textsc{Small Steps for Analytics} and to incrementally achieve the \textsc{Reachable Goals}, being helpful especially when the tasks related to implementing improvements defined in the analytics process are neglected in planning and frequently left out of the iterations.

\section{Broad-spectrum Diagnostic}

Introducing Software Analytics to software teams can be challenging, especially when those teams have low experience with analytics practices and haven't adopted any in their projects. They can have trouble understanding the extent to which they would benefit from such practices, or the kind of analytics results that one can learn from. This can make them abandon the use of analytics at the start or not correctly use Software Analytics.

\subsection{Problem}

\textbf{The team doesn't have an understanding of the type of questions that can be answered through software analytics.}
\vspace{0.5em}

Even though Software Analytics practices help solve code quality issues and improve software quality based on code metrics and insights of those metrics, teams that have no experience with these practices often have difficulties understanding what issues can be solved, verified or improved with the help of metrics and their analysis. This can make it hard for a team to figure our our \textsc{What You Need to Know} as it might be clear to them the kind of things that \textit{can} be known. This may make teams stuck at the very start of trying to adopt Software Analytics practices. It can make teams can lose interest and their perceptions of the usefulness of these practices can decrease which can make them abandon the use of analytics in their development process.

\subsection{Solution}

\textbf{Use tools that make a general diagnostic of the system and show a couple of metrics, to help detect a set of initial issues that can be used as a starting point to the software analytics practices' usage.}
\vspace{0.5em}

Start analytics adoption by providing team members with some examples of metrics and issues that can be solved or verified in their projects through Software Analytics. Firstly, choose a general area of concern in which the team believes it might be beneficial to have the application evaluated, such as code quality, test coverage or runtime execution. Secondly, choose a tool that can collect information and perform a more general assessment of the system on the desired area.

For instance, to have a vision of the overall quality of the system, a static analytics tool, such as SonarQube, can detect potential problems and provide a few metrics, like cognitive and cyclomatic complexity, code duplication, deprecated methods use, or possible security vulnerabilities. Additionally, dynamic analysis tools can also be of use, for example, some cloud providers provide services to collect data from the application runtime environments and provide tools to access and inspect such information. 

By looking at this more general and broad diagnostic performed by the default configuration of such tools, the team will gain a practical understanding of what issues can be solved and what insights can be gained from metrics. This analysis should be taken carefully because some general-purpose metrics tools shows the same kind of metrics to every project, so it will be up to the team to decide which metrics and issues are more helpful within their context. Many tools give severity levels to the issues they identify, and looking at the most critical ones can be a good starting point.

After improving the understanding of what metrics can be obtained, and the kind of issues that they allow to identify, developers will be in a better position to figure out \textsc{What you need to know}
Having in mind some of the issues found in this general analysis, the team may now be able to follow the rest of the analytics process, and further improve their understanding how to use its practices, and the patterns in Section~\ref{sec:patters_sa} can guide them to solve different kinds of problems.  

\subsection{Consequences}

As a consequence, the team members can learn more about Software Analytics and better understand which metrics they can adopt as part of a measurement plan to handle the issues raised as significant. Additionally, it also helps the team to understand how to use the analytics as part of their development and improvement process. In the end, the team will have already a tool integrated with the environment which can be useful in the future because some metrics are already available to analyze. This tool can be configured and tuned for more specific analysis that will fit better in the project interests. 

A negative consequence is that the tool's report can guide the team in the wrong direction because without a good analysis of the tool's results they can be looking to metrics and issues that are not relevant for their project or cause the team to lose interest or lead them to wrong conclusions based on irrelevant metrics.

\subsection{Related Patterns}

\textsc{Continuous Inspection}~\cite{merson2014continuous} is about how to detect architecture and code problems as soon as possible by prescribing the use available automated tools to continuously inspect code, generate a report on the overall code health. The analysis performed in the \textsc{Broad-spectrum Diagnostic} should not be included by default in the \textsc{Continuous Inspection}, and that can be done after to \textsc{Define Quality Standards}.

\textsc{System Quality Dashboards}~\cite{yoder2014qa} is useful to show real-time results and display quality values measured during check-in or system build quality tests. Since the goal of \textsc{Broad-spectrum Diagnostic} is to be a starting point, its results should not be integrated by default into this kind of dashboard.

After having a broad view of potential issues through the \textsc{Broad-spectrum Diagnostic}, the next step would be to define \textsc{What You Need to Know}~\cite{choma2017patterns} to focus on the most pressing issues and those that will add the most value to the project.

\subsection{Known Uses}

\begin{itemize}
    \item We have ourselves applied the pattern in an industrial software project. The context was that of a recent project within a large company, with a team of 3 developers and various clients governing the project and giving feedback. We understood that the team did not have experience with software analytics. Most decisions were not data-driven, they were based on personal experience and on feedback from product owners and clients.

We have introduced the Software Analytics practices to the team, but team members struggled to understand the objectives and issues that could be solved using analytics. Namely, on our first try they did not come to any conclusion regarding \textsc{What you need to know}.

We then decided to use another approach. We started using a metrics tool---SonarQube---and showed the team its report, with all the issues it pointed out and some interesting metrics to promote discussion. By analyzing the importance of each of these issues and metrics, it was easier to understand what could be found by the use of software metrics and start a discussion on \textsc{What you need to know} given the specific context of this project. With this approach, we were able to move forward, and the team started to bring more issues to the table and understand better the objectives of Software Analytics.

\item To identify success factors that help teams to create better deliveries in future releases and failure factors that help teams to prevent bad deliveries, \citeN{huijgens2017strong} carried out in an exploratory study in an international bank with more than 300 teams and about 750 different applications. In this study, they defined a limited set of software metrics focused on a delivery scope (\textit{e.g.}, epics, user stories). However, to define the most relevant lagging metrics and related strong leading metrics they need to explore other data sources.

\item In a project conducted in a Brazilian company, the team already had SonarQube installed in their environment but did not use it regularly. When asked about relevant issues to be handled by the analytics process, the team did not know precisely what kind of issue they could investigate. However, by looking at the result generated by default by SonarQube, the team was able to quickly identify some issues to be further investigated. The information provided by the tool was not enough to solve the issues, but it was important for the team to understand the kind of issues that could be interesting to address. 

\end{itemize}

\section{Embedded Improvements}
{\small Also know as \textit{Embed Improvements in Tasks}.}\\

Improvement tasks that emerge from the use of software analytics are often large and difficult to estimate. This type of task can often appear daunting to start and can be left untouched in the backlog. In legacy systems, this type of issue might not even be worth fixing since the codebase is old or the quality has not been the main focus, and most developers might not have worked on that specific part of the system.

\subsection{Problem}

\textbf{Software analytics practices often generate technical improvement tasks that imply a considerable effort, and therefore get frequently left behind during planning and may never end up being implemented.}
\vspace{0.5em}

There usually is a lack of understanding from the end-users and product owners, which are responsible for the product direction, regarding code-quality issues. They may not be able to identify internal quality concerns and understand the importance of dedicating some effort to addressing them. Taking on considerably large technical improvements may not allow to, in each iteration, deliver the desired quantity of business value. 

Moreover, iteration planning often favors tasks with a clear business value, and other kind of improvements may remain in product backlogs, postponing improvements that may be much-needed but difficult to prioritize by non-technical stakeholders.

Additionally, there can be a significant effort in some tasks derived from software analytics findings. The team can try to split such tasks into smaller ones and implement them over time with Small Steps  for Analytics. Although this might be effective, it might not be enough if future developments make the same issue resurface and, sometimes, the issue might simply be too complex to be divided into smaller tasks. The task is indeed being resolved, but this doesn't necessarily imply any plans to prevent the issue from happening.

\subsection{Solution}

\textbf{Change the development practices to embed improvements gradually and continuously in other tasks.}
\vspace{0.5em}

Changing the development practices is essential for this pattern to work, and there are two complementary ways a team can adopt this pattern.

This first way is when a team changes their DoD (Definition of Done)~\cite{dod} to prevent an issue from happening again. The DoD is a checklist that needs to be completely done before considering a task as completed. The team will have the responsibility to make sure that the issue is not happening before closing a task. This way, the issue does not need to be brought up during the planning. It will simply be an underlying requirement when doing a specific task. The team must be aware that the issue will still be present in what was previously done and should consider if it also needs to be resolved. As an example, the DoD might require a minimum value for a test coverage metric, if the goal is to improve the tests.

Another way is to define guidelines to be followed by new features and perform refactoring before changing an existing code to introduce the proposed guideline. That will prevent the problem from appearing in new code, and at a small pace, will migrate the existing one to the new approach. That also avoids having an enormous refactoring task only focused on performing the change. To exemplify this alternative, if the team defined to migrate to a new library, refactoring can be added as a task from a User Story that would require a change in this code. Moreover, any new development should be done using the new library.

\subsection{Consequences}

There are several consequences when using this pattern. On the positive side, is the guarantee that the problem won't resurface. Considering the issue on new developments or including a specific check on the DoD, prevents a particular issue from happening again.

Another positive consequence is that there is no need to create additional tasks to fix or improve an issue. As stated before, tasks related to significant issues can be daunting to start. Considering the improvement as part of typical developments, it won't feel like a substantial endeavor, and progress, although small, will always be made towards the end goal. On the other hand it depends on a disciplined team, that will be sure to follow the DoD established by the team.

On the negative side of the consequences, time management is one of the main problems. The team might take more time to finish a task since there are more things to consider. It may reduce the team's velocity on the short run, and the product owners and clients might notice the change. On the other hand, it may increase the team's velocity on the long run, as the improved code quality may allows future developments to be quicker to implement. The team must evaluate if refactoring makes sense at any particular stage, and contract technical debt if that is indeed the better option.

Another consequence that can arise from subsequent uses of this solution is the cluttering of the \textit{DoD}. It might start to be cluttered with small steps to prevent multiple issues that stemmed from the use of the SA Canvas. Although, in a way, this is positive for the system quality, for the developer might be hard to consider everything before closing a task or User Story. To avoid that, these Embedded Improvements should be incorporated by the team naturally and become part of their day-by-day practices.

An additional consequence is that although this prevents the issue from happening again, the issue persists on older code and needs to be taken care of to eliminate the system's issue. If a team wants to eliminate the process, they have to create tasks to address the issue on older code incrementally, which is still a problem if the issue is large and complex.

\subsection{Related Patterns}

\textsc{Integrate Quality}~\cite{yoder2014} is about how incorporate quality assurance into software process by including a lightweight means for describing and understanding system qualities. The use of \textsc{Embedded Improvements} is an approach to implement that, but it is not the only one, since \textsc{Integrate Quality} is more general.

\textsc{Reachable Goals}~\cite{choma2017patterns} pattern helps teams identify achievable goals before planning and implementing their actions. While this pattern focus on how an approach to establish objectives for the team, \textsc{Embedded Improvements} propose an approach on how to achieve them. In this pattern language, \textsc{Small Steps for Analytics}~\cite{choma2017patterns} is also related to this one, however \textsc{Embedded Improvements} is a more specific solution in that direction.

\textsc{Quality Stories}~\cite{yoder2014} recommends creating stories that specifically focus on some measurable quality issues of the system that must be achieved, which can be useful to the team for prioritizing and including these quality items on the backlog. That can be considered a competing approach to  \textsc{Embedded Improvements}, since it proposes to hide these activities embedding them into existing tasks.

\subsection{Known Uses}

\begin{itemize}
\item We have ourselves applied the pattern in an industrial software project. The context was that of a start-up that had come to accumulate significant technical debt and was dealing with different technical challenges. Fixing some of the issues that the teams identified were considerably large endeavors.

We found \textsc{Embedded improvements} useful in the context of this start-up in more than one occasion. The teams were challenged with many issues that they already knew existed but still hadn't formally organized and decided how to address. Some of these issues needed considerable time to resolve. One of them was related to dead code or code that wasn't being used anymore. With a simple script that analyzed the server logs, the team managed to find out that 70\% of their core application endpoints were not being called anymore. Removing a large portion of code like this implies a few risks, so tackling it in smaller segments was one of the solutions that was put into practice (\textit{i.e.}, \textsc{Small Steps for Analytics}). But the team also wanted to prevent the issue from reappearing, so they started analyzing the possibility of generating dead code in new refinements. If this was the case, the portion of code in question should be removed to prevent leaving dead code in the repository.

Another use of the pattern is related to test coverage. Some older projects had low coverage but adding tests to all components that were still missing them was very hard, since the code had been done years ago, sometimes by developers that had since left the company. Since creating all these tests was not an option, the team set a goal that all new code should have 100\% coverage. This decision is expected to make code coverage increase over time.

\item \citeN{snyder2018using} reported how software analytics were used to guide improvements and evaluate progress during an Agile and DevOps transformation in a software company. They reported that to detect structural-quality flaws and produce the analytic measures, the teams began scanning their builds at a minimum of once per sprint (every two weeks), enabling them to address the most critical issues before release. By scanning several times a week or even daily, the team could fix critical defects in a day or two, rather than waiting until the next sprint. 

\item The lack of tests, evidenced by a low code coverage metric, was identified as a problem by a development team of a Brazilian company. To address it, the team decided to create more tests for some specific classes, and incorporated a new task in the backlog with such objective. However this task remained in the backlog after a few planning sessions, and was never considered to have enough priority for inclusion in one of the iterations. A reason for this is that it was a task that required considerable effort to be completed. The team then decided to embed the improvement of test coverage in other user stories---the code created or changed in the context of each user story would have to have the desired code coverage. The desired coverage for the system as a whole was not reached immediately, but the code coverage started finally to improve in next iterations.

\end{itemize}

\section{Summary}
This paper presented two new patterns to compose the Pattern Language for Software Analytics. The complete set of patterns includes ways of incorporating software analytics activities within software development projects.The pattern \textsc{Broad-Spectrum Diagnostics} was identified for the scenario that we faced when the team does not understand the kind of problem that they can solve using analytics, \textit{a priori}. While, the pattern \textsc{Embedded Improvements} refers to embed improvements gradually and continuously by adopting a checklist of tasks to ensure they are done or considering the issues in engineering refinements as a new task or user story.

\section{Acknowledgements}
The authors would like to thank Uwe Zdun who graciously shepherded this paper for PLoP 2021, to Justus Bogner who provided insights on earlier versions of the work and, finally, to all the participants of the the Napa writers' workshop at PLoP 2021---Richard Gabriel, Joe Yoder, Jonathan Edwards, Christopher Hartley, Tomas Petricek, Michael Weiss and Daniel Pinho. Thank you for generously contributing to the paper through discussion, and by providing insightful feedback that helped us to greatly improve our work.

We would also like to thank the companies who we have worked with, and that made it possible for us to identify these patterns, as well as the Integrated Masters in Informatics Engineering of the Faculty of Engineering of the University of Porto, for supporting this work. 

\bibliographystyle{ACM-Reference-Format-Journals}
\bibliography{main}


\begin{thebibliography}{00}


\ifx \showCODEN    \undefined \def \showCODEN     #1{\unskip}     \fi
\ifx \showDOI      \undefined \def \showDOI       #1{{\tt DOI:}\penalty0{#1}\ }
  \fi
\ifx \showISBNx    \undefined \def \showISBNx     #1{\unskip}     \fi
\ifx \showISBNxiii \undefined \def \showISBNxiii  #1{\unskip}     \fi
\ifx \showISSN     \undefined \def \showISSN      #1{\unskip}     \fi
\ifx \showLCCN     \undefined \def \showLCCN      #1{\unskip}     \fi
\ifx \shownote     \undefined \def \shownote      #1{#1}          \fi
\ifx \showarticletitle \undefined \def \showarticletitle #1{#1}   \fi
\ifx \showURL      \undefined \def \showURL       #1{#1}          \fi

\bibitem[\protect\citeauthoryear{Augustine, Hudepohl, Marcinczak, and
  Snipes}{Augustine et~al\mbox{.}}{2018}]%
        {augustine2018deploying}
{Vinay Augustine}, {John Hudepohl}, {Przemyslaw Marcinczak}, {and} {Will
  Snipes}. 2018.
\newblock \showarticletitle{Deploying Software Team Analytics in a
  Multinational Organization}.
\newblock {\em IEEE Software\/} {35}, 1 (2018), 72--76.
\newblock


\bibitem[\protect\citeauthoryear{Buse and Zimmermann}{Buse and
  Zimmermann}{2010}]%
        {buse2010analytics}
{Raymond~PL Buse} {and} {Thomas Zimmermann}. 2010.
\newblock \showarticletitle{Analytics for software development}. In {\em
  Proceedings of the FSE/SDP workshop on Future of software engineering
  research}. ACM, Santa Fe, New Mexico, USA, 77--80.
\newblock


\bibitem[\protect\citeauthoryear{Choma, Guerra, da~Silva, Zaina, and
  Correia}{Choma et~al\mbox{.}}{2019}]%
        {chomatowards}
{Joelma Choma}, {Eduardo~M Guerra}, {Tiago~Silva da Silva}, {Luciana~AM Zaina},
  {and} {Filipe~Figueiredo Correia}. 2019.
\newblock \showarticletitle{Towards an artifact to support agile teams in
  software analytics activities}. In {\em Proceedings of the International
  Conference on Software Engineering and Knowledge Engineering (SEKE)}. KSI
  Research Inc, Lisbon, Portugal, 88--93.
\newblock


\bibitem[\protect\citeauthoryear{Choma, Guerra, and Silva}{Choma
  et~al\mbox{.}}{2017}]%
        {choma2017patterns}
{Joelma Choma}, {Eduardo~M Guerra}, {and} {Tiago~S Silva}. 2017.
\newblock \showarticletitle{Patterns for Implementing Software Analytics in
  Development Teams}. In {\em Proceedings of the 24th Conference on Pattern
  Languages of Programs}. ACM, Vancouver, Canada, 12.
\newblock


\bibitem[\protect\citeauthoryear{Choma, Guerra, and Silva}{Choma
  et~al\mbox{.}}{2018}]%
        {choma2018patterns}
{Joelma Choma}, {Eduardo~M Guerra}, {and} {Tiago~S Silva}. 2018.
\newblock \showarticletitle{Learning from Experiments, Define Quality
  Standards, Suspend Measurement: Three patterns in a Software Analytics
  Pattern Language}. In {\em Proceedings of the 12th Latin American Conference
  on Pattern Languages of Programs (SugarLoafPLoP)}. ACM, Valparaíso, Chile,
  10.
\newblock


\bibitem[\protect\citeauthoryear{Davenport}{Davenport}{2009}]%
        {davenport2009make}
{Thomas~H Davenport}. 2009.
\newblock \showarticletitle{Make better decisions}.
\newblock {\em Harvard business review\/} {87}, 11 (2009), 117--123.
\newblock


\bibitem[\protect\citeauthoryear{Davenport}{Davenport}{2010}]%
        {davenport2010organizations}
{Thomas~H Davenport}. 2010.
\newblock {\em How organizations make better decisions}.
\newblock {T}echnical {R}eport. International Institute for Analytics.
\newblock


\bibitem[\protect\citeauthoryear{Hassan and Xie}{Hassan and Xie}{2010}]%
        {hassan2010software}
{Ahmed~E Hassan} {and} {Tao Xie}. 2010.
\newblock \showarticletitle{Software intelligence: the future of mining
  software engineering data}. In {\em Proceedings of the FSE/SDP workshop on
  Future of software engineering research}. ACM, Santa Fe, New Mexico, USA,
  161--166.
\newblock


\bibitem[\protect\citeauthoryear{Huijgens, Lamping, Stevens, Rothengatter,
  Gousios, and Romano}{Huijgens et~al\mbox{.}}{2017}]%
        {huijgens2017strong}
{Hennie Huijgens}, {Robert Lamping}, {Dick Stevens}, {Hartger Rothengatter},
  {Georgios Gousios}, {and} {Daniele Romano}. 2017.
\newblock \showarticletitle{Strong Agile Metrics: Mining Log Data to Determine
  Predictive Power of Software Metrics for Continuous Delivery Teams}. In {\em
  Proceedings of the 2017 11th Joint Meeting on Foundations of Software
  Engineering} {\em (ESEC/FSE 2017)}. Association for Computing Machinery, New
  York, NY, USA, 866–871.
\newblock
\showISBNx{9781450351058}
\showDOI{%
\url{http://dx.doi.org/10.1145/3106237.3117779}}


\bibitem[\protect\citeauthoryear{Huijgens, Spadini, Stevens, Visser, and van
  Deursen}{Huijgens et~al\mbox{.}}{2018}]%
        {huijgens2018software}
{Hennie Huijgens}, {Davide Spadini}, {Dick Stevens}, {Niels Visser}, {and}
  {Arie van Deursen}. 2018.
\newblock \showarticletitle{Software analytics in continuous delivery: a case
  study on success factors}. In {\em Proceedings of the 12th ACM/IEEE
  International Symposium on Empirical Software Engineering and Measurement}.
  ACM, Oulu, Finland, 25.
\newblock


\bibitem[\protect\citeauthoryear{Madan}{Madan}{2019}]%
        {dod}
{Sumeet Madan}. 2019.
\newblock DONE Understanding Of The Definition Of "Done".
\newblock   (Dec 2019).
\newblock
\showURL{%
\url{https://www.scrum.org/resources/blog/done-understanding-definition-done}}


\bibitem[\protect\citeauthoryear{Merson, Aguiar, Guerra, and Yoder}{Merson
  et~al\mbox{.}}{2013}]%
        {merson2014continuous}
{P Merson}, {A Aguiar}, {E Guerra}, {and} {J Yoder}. 2013.
\newblock \showarticletitle{Continuous inspection: a pattern for keeping your
  code healthy and aligned to the architecture}. In {\em Proceedings of the 2nd
  Asian Conference on Pattern Languages of Programs (AsianPLoP).} ACM, Tokyo,
  Japan, 6--8.
\newblock


\bibitem[\protect\citeauthoryear{Robbes, Vidal, and Bastarrica}{Robbes
  et~al\mbox{.}}{2013}]%
        {robbes2013software}
{Romain Robbes}, {René Vidal}, {and} {María~Cecilia Bastarrica}. 2013.
\newblock \showarticletitle{Are Software Analytics Efforts Worthwhile for Small
  Companies? The Case of Amisoft}.
\newblock {\em IEEE Software\/} {30}, 5 (2013), 46--53.
\newblock


\bibitem[\protect\citeauthoryear{Schmitt, Rosing, Zhang, and
  Leatherbee}{Schmitt et~al\mbox{.}}{2018}]%
        {startupuncertainties}
{Antje Schmitt}, {Kathrin Rosing}, {Stephen~X Zhang}, {and} {Michael
  Leatherbee}. 2018.
\newblock \showarticletitle{A dynamic model of entrepreneurial uncertainty and
  business opportunity identification: Exploration as a mediator and
  entrepreneurial self-efficacy as a moderator}.
\newblock {\em Entrepreneurship Theory and Practice\/} {42}, 6 (2018),
  835--859.
\newblock


\bibitem[\protect\citeauthoryear{Snyder and Curtis}{Snyder and Curtis}{2018}]%
        {snyder2018using}
{Barry Snyder} {and} {Bill Curtis}. 2018.
\newblock \showarticletitle{Using Analytics to Guide Improvement during an
  Agile--DevOps Transformation}.
\newblock {\em IEEE Software\/} {35}, 1 (2018), 78--83.
\newblock


\bibitem[\protect\citeauthoryear{Storey}{Storey}{2016}]%
        {storey2016lies}
{M-A Storey}. 2016.
\newblock \showarticletitle{Lies, damned lies, and analytics: Why big data
  needs thick data}.
\newblock In {\em Perspectives on Data Science for Software Engineering}, {Tim
  Menzies}, {Laurie Williams}, {and} {Thomas Zimmermann} (Eds.). Morgan
  Kaufmann, Boston, 369--374.
\newblock


\bibitem[\protect\citeauthoryear{Yoder and Wirfs-Brock}{Yoder and
  Wirfs-Brock}{2014}]%
        {yoder2014qa}
{Joseph~W Yoder} {and} {Rebecca Wirfs-Brock}. 2014.
\newblock \showarticletitle{QA to AQ part two: shifting from quality assurance
  to agile quality:" measuring and monitoring quality"}. In {\em Proceedings of
  the 21st Conference on Pattern Languages of Programs}. ACM, Monticello, IL,
  USA, 1--20.
\newblock


\bibitem[\protect\citeauthoryear{Yoder, Wirfs-Brock, and Aguiar}{Yoder
  et~al\mbox{.}}{2014}]%
        {yoder2014}
{Joseph~W Yoder}, {Rebecca Wirfs-Brock}, {and} {Ademar Aguiar}. 2014.
\newblock \showarticletitle{QA to AQ: Patterns about transitioning from Quality
  Assurance to Agile Quality}. In {\em Proceedings of the 3rd Asian Conference
  on Pattern Languages of Programs, Tokyo, Japan}. ACM, Monticello, IL, USA,
  12.
\newblock


\bibitem[\protect\citeauthoryear{Zhang, Dang, Lou, Han, Zhang, and Xie}{Zhang
  et~al\mbox{.}}{2011}]%
        {zhang2011software}
{Dongmei Zhang}, {Yingnong Dang}, {Jian-Guang Lou}, {Shi Han}, {Haidong Zhang},
  {and} {Tao Xie}. 2011.
\newblock \showarticletitle{Software analytics as a learning case in practice:
  Approaches and experiences}. In {\em Proceedings of the International
  Workshop on Machine Learning Technologies in Software Engineering}. ACM,
  Lawrence, Kansas, USA, 55--58.
\newblock


\bibitem[\protect\citeauthoryear{Zhang, Han, Dang, Lou, Zhang, and Xie}{Zhang
  et~al\mbox{.}}{2013}]%
        {zhang2013software}
{Dongmei Zhang}, {Shi Han}, {Yingnong Dang}, {Jian-Guang Lou}, {Haidong Zhang},
  {and} {Tao Xie}. 2013.
\newblock \showarticletitle{Software analytics in practice}.
\newblock {\em IEEE software\/} {30}, 5 (2013), 30--37.
\newblock


\end{thebibliography}

\received{June 2021}{September 2021}{February 2022}

\end{document}